\documentclass[aps,prl,twocolumn,showpacs,superscriptaddress]{revtex4-1}

\usepackage{amsfonts,mathrsfs,amsmath,amsthm,amssymb,bbold}
\usepackage{epsfig}
\usepackage{bm}
\usepackage{xfrac}
\usepackage{tabulary}
\usepackage{color}
\usepackage[dvipsnames]{xcolor}

\begin{document}  
\title {\bf Modular-value-based metrology with spin coherent pointers}

\author{ Le Bin Ho}
\thanks{Electronic address: binho@kindai.ac.jp}
\affiliation{Department of Physics, Kindai University, Higashi-Osaka, 577-8502, Japan}

\author{Yasushi Kondo}
\affiliation{Department of Physics, Kindai University, Higashi-Osaka, 577-8502, Japan}
\affiliation{Interdisciplinary Graduate School of Science and Engineering, 
Kindai University, Higashi-Osaka, 577-8502, Japan}

\date{\today}

\begin{abstract}

Modular values are quantities that described by pre- and postselected states of 
quantum systems like weak values but are different from them: The associated 
interaction is not necessary to be weak. We discuss an optimal 
modular-value-based measurement with a spin coherent pointer:
A quantum system is exposed to a field in which strength is to be estimated 
through its modular value.
We consider two cases, with a two-dimensional and a higher-dimensional 
pointer, and evaluate the quantum Fisher information.
The modular-value-based measurement has no merit in the former case, 
while its sensitivity can be enhanced in the latter case. 
We also consider the pointer under a phase-flip error.
Our study should motivate researchers to apply the modular-value-based 
measurements for quantum metrology.

\end{abstract}
\maketitle

{\em Introduction.}---In the concept of quantum sensing, a physical quantity 
on a small scale can be measured indirectly via a quantum system, 
a quantum property, or a quantum phenomenon \cite{Degen89}. 
The principle of measurements is (i)~preparing 
quantum systems, hereafter called {\it sensor}s, of which number is $L$, 
(ii)~exposing to a field of which strength is to be measured 
for a period of $t$, and (iii)~obtaining 
a state change before and after the exposure. The change is a measure of 
the strength of the field. 
This procedure is repeated $T/t$ times in a total measurement time $T$. If the 
sensors are independent, the effective total measurement number $N$ is given as 
$N = LT/t$. For fixed $T$ and $t$, the uncertainty of the estimation is proportional 
to $1/\sqrt{L}$, which is known as the standard quantum limit or the shot noise limit 
\cite{Matsuzaki84,Pezze102,Huelga79,Itano47}. However, if the uncertainty scales as  
$1/L$ then it is called the Heisenberg limit \cite{Zwierz105} 
which is a fundamental limit. 

There have been a number of attempts to improve measurement protocols to overcome 
the standard quantum limit.
One of the pioneering attempts is to employ squeezed states to reduce noises 
\cite{Wineland46,Wineland50,Giovannetti306}. 
The entanglement is also a resource for defeating the standard quantum limit 
\cite{Giovannetti306,Giovannetti96,Huelga79,Pezze102,Jones342,Simmons82}. 
For example, Jones {\it et al.} have used $\it NOON$ states 
\cite{Jones342,Simmons82}. 
Zaiser {\it et al.} have claimed that the sensitivity 
can be significantly enhanced  by using a quantum memory \cite{Zaiser7}. 
Matsuzaki {\it et al.} have proposed a protocol with a teleportation 
\cite{Matsuzaki120}.

These studies focused on sensors exposed to a field.  
We, here,  draw attention a method how the sensors are measured. 
One approach in this 
direction is employing a weak value \cite{Aha60,Aha2005,Kofman520,Dressel86}. 
Measurements with weak values had been expected to enhance the sensitivity 
\cite{Alves95,Zhang114,Alves91,Tanaka88,Knee87,Knee4}, but it turned out 
that they can only reach the standard quantum limit 
\cite{Alves95,Zhang114,Alves91,Tanaka88,Knee87,Knee4}.
The measurements with weak values, however, can be dramatically improved 
when sensors are entangled \cite{Pang113,Pang92}. Furthermore, 
by employing non-classical pointer states, 
the sensitivity can reach the Heisenberg limit \cite{Zhang114,Pang115, Jordan2}.
 
In this Letter, we discuss modular-value-based measurements 
\cite{Kedem105,Ho380,Ho95,Ho59,Ho97}
with spin-$j$ coherent pointers \cite{Amiet24,Fabian93}. 
They are different from the weak-value-based ones and can allow arbitrary 
strength interactions between the pointers and sensors. 
The pointer may be considered as a {\it measurement device} for extracting 
the field information from the sensors.  
In order to evaluate the sensitivity, we focus on the quantum Fisher information 
contained in the Cram\'er-Rao inequality 
\cite{Zwierz105,Hofmann86,Braunstein72,Rao}. 
Its maximum provides the lower 
bound on the sensitivity for measuring the field. 
In this Letter, 
$L$ is fixed and we compare the Fisher 
information of the measurements under various conditions. 
We first examine a measurement with a $j=\frac{1}{2}$ 
(qubit) pointer and then move our attention 
to that with a $j\ge 1$ pointer. 
We also consider the case when the qubit pointer is 
under a phase-flip error \cite{Nielsen}.

{\em Modular-value-based measurement.}---
We assume that 
the sensor exposed to a field is a qubit 
and that its initial state  is 
$|\psi_i\rangle = \frac{1}{\sqrt{2}}(|0\rangle+|1\rangle)$
which evolves to 
$|\psi_i(\omega)\rangle = 
\frac{1}{\sqrt{2}}(|0\rangle + e^{-i\omega t}|1\rangle)$ 
during a period $t$. $\omega$ is a measure of the field
strength to be {\it estimated}. 
We prepare the pointer in a spin-$j$ coherent state
\begin{align}
\label{spinjc}
|\xi\rangle = 
\sum_{m=-j}^j\sqrt{{2j \choose j+m}}\Bigl(\cos\frac{\theta}{2}\Bigr)^{j+m}
\Bigl(\sin\frac{\theta}{2}e^{-i\varphi}\Bigr)^{j-m}|j,m\rangle
\end{align}
in spherical representation  \cite{Amiet24,Fabian93},
where $\theta \in [0,\pi]$ and $\varphi \in [0, 2\pi]$.  
$|j,m\rangle$ is the standard angular momentum
basis, for a fixed $j$, $m\in [-j, +j]$. 

The interaction Hamiltonian between the sensor and the pointer 
is assumed to be $\hat{H} = g(\tau){\sigma}_z\otimes |j,+j\rangle\langle j,+j|$. 
After the interaction during $\tau = 0 \sim \tau_0$, the joint state of 
the sensor-pointer yields
\begin{align}\label{sys_poin_stat_aft}
|\Psi(\omega)\rangle = \dfrac{1}{\sqrt{2}}\Bigl(|0\rangle\otimes|\xi_+\rangle 
+ e^{-i\omega t}|1\rangle\otimes|\xi_-\rangle\Bigr),
\end{align}
where $|\xi_\pm\rangle = \exp(\mp ig|j,+j\rangle\langle j,+j|)|\xi\rangle$
and $g=\int_0^{\tau_0} g(\tau)d\tau$ is the coupling strength. 
After postselecting the sensor onto a final state 
$|\psi_f\rangle = \cos\frac{\phi}{2}|0\rangle+\sin\frac{\phi}{2}|1\rangle$, 
the normalized pointer state becomes 
\begin{align}\label{pointer_final}
\notag & |\eta\rangle\\
\notag &= \langle\psi_f|\hat{U}|\psi_i(\omega)\rangle\otimes|
\xi\rangle/\sqrt{p(\omega)}\\
&= \frac{\langle\psi_f|\psi_i(\omega)\rangle\Bigl[\sum_{m=-j}^{j-1}c_{jm} |j,m\rangle 
+ \Bigl(\cos\frac{\theta}{2}\Bigr)^{2j}({\sigma}_z)_{\rm m}|j,j\rangle\Bigr]}
{\sqrt{p(\omega)}}.
\end{align}
Here, $p(\omega) = ||\langle\psi_f|\hat{U}|\psi_i(\omega)\rangle\otimes|
\xi\rangle||^2$ 
is the probability of successful postselection, where $||*||$ is the {\it norm} of ($*$),

$c_{jm}=\sqrt{{2j \choose j+m}}\bigl(\cos\frac{\theta}{2}\bigr)^{j+m}
\bigl(\sin\frac{\theta}{2}e^{-i\varphi}\bigr)^{j-m}$, and 
\begin{eqnarray*}
({\sigma}_z)_{\rm m} &=& 
\frac{\langle\psi_f|e^{-ig{\sigma}_z}|\psi_i(\omega)\rangle}
{\langle\psi_f|\psi_i(\omega)\rangle} 
\end{eqnarray*}
is $\omega $ dependent modular value of the observable ${\sigma}_z$. 
Note that $({\sigma}_z)_{\rm m}$ is independent of the choice of the initial pointer 
state $|\xi\rangle$. 
Throughout this Letter, we fix $g=\pi/2$. Under this condition, 
the modulus of the modular value, 
$|({\sigma}_z)_{\rm m}|$, becomes the minimum 
when $|\psi_f\rangle = |\psi_i\rangle$ ($\phi = \pi/2$)
and does the maximum 
when $|\psi_f\rangle$ is orthogonal to $|\psi_i\rangle $ ($\phi = 3\pi/2$).

The uncertainty $\delta \omega $ in the estimation of $\omega$ after $N$ 
independent measurements is defined by 
$\displaystyle \delta\omega = 
\sqrt{\langle(\omega- \overline{\omega})^2\rangle} $, 
where $\overline{\omega}$ is the average of the measured $\omega$'s.
Its minimum is determined by the Cram\'er-Rao lower bound as 
$\delta\omega \geq 1/\sqrt{NF}$ 
\cite{Zwierz105,Hofmann86,Braunstein72,Rao}. 
$F$ is the Fisher information and is defined by 
$F = \sum_k[1/p_k(\omega)][dp_k(\omega)/d\omega]^2$,
where $p_k(\omega)$ is the probability distribution for obtaining 
the $k$'th experimental result. The maximization of $F$ 
over all possible measurements 
leads to the quantum Fisher information 
$Q$ \cite{Helstrom,Holevo,Braunstein72}. 
For a pure quantum state $|\psi(\omega)\rangle$, $Q$ 
is given by \cite{Helstrom}
\begin{align}\label{Qfisher}
Q= 4\Bigl[\dfrac{d\langle\psi(\omega)|}
{d\omega}\dfrac{d|\psi(\omega)\rangle}{d\omega}-\Bigl|
\dfrac{d\langle\psi(\omega)|}{d\omega}|\psi(\omega)\rangle\Bigr|^2\Bigl].
\end{align}
In order to obtain the best sensitivity, we have to maximize $Q$
when $N$ is fixed.

We define the quantum Fisher information of the sensor of which state is 
$|\psi_i(\omega)\rangle$ as $Q_0$. 
By following Eq.~(\ref{Qfisher}), we obtain $Q_0 = t^2$. 
$Q_0$ is also obtained in the case of Ramsey sensing 
\cite{Matsuzaki84,Huelga79}. Therefore, we call these measurement 
protocols which give $Q = t^2$ as the conventional measurement. 
$Q_0$ is employed as the reference when comparing the Fisher information 
in various measurements in this Letter. 

In measurement protocols with postselection,  $Q$ 
is given by the sum of  $Q_m$ and 
$F_p$ 
\cite{Zhang114,Combes89,Alves91,Alves95}.
$Q_m$ is called the {\em measured quantum Fisher information} 
which is a product of the Fisher information of the final pointer state 
and the probability of successful postselection $p(\omega)$.  
$F_p$ is called the {\em postselected classical Fisher information} 
defined as  \cite{Alves91,Alves95}
\begin{align}\label{Fisher_p}
F_p= \dfrac{1}{p(\omega)[1-p(\omega)]}
\Bigl[\dfrac{dp(\omega)}{d\omega}\Bigr]^2.
\end{align}
As shown in the Supplementary Material \cite{Sub},
$F_p = O(\omega^2)$ at $\omega t \ll 1$
for all $j$. Therefore, we can ignore $F_p$ and 
we have to consider only  $Q_m$. 
We rewrite the final 
pointer state Eq.~\eqref{pointer_final} as
\begin{align}\label{pointer_final_conv}
|\eta\rangle = 
\dfrac{1}{\sqrt{2}}\Bigl(\cos\frac{\phi}{2}|\xi_+\rangle+
\sin\frac{\phi}{2}e^{-i\omega t}|\xi_-\rangle\Bigr)/\sqrt{p(\omega)}
\end{align}
and then substitute it into 
Eq.~\eqref{Qfisher}.
We will evaluate $Q_m$ in two cases 
of $j = 1/2$ and $j \ge 1$. 
We introduce  $Q^{(j)}_m$ which denotes 
the measured quantum Fisher information 
with the spin-$j$ coherent pointer in our measurement protocol. 

\begin{figure} [h]
\includegraphics[width=8.5cm]{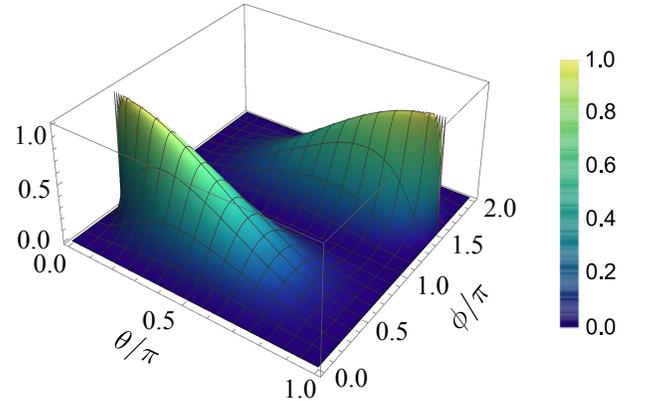}
\caption{
(Color online) 
$Q^{(\sfrac{1}{2})}_m/Q_0$
as a function of $\theta$ and $\phi$.  
}
\label{Fmq}
\end{figure}

{\em Two-dimensional pointer.}---We evaluate $Q^{(\sfrac{1}{2})}_m$.
According to the definition of $Q_m$, we substitute 
Eq.~\eqref{pointer_final_conv} into Eq.~\eqref{Qfisher} and make a product 
with $p(\omega)$ \cite{Sub}, we obtain
\begin{align}\label{fisher_modular}
Q^{(\sfrac{1}{2})}_m
= \dfrac{t^2\sin^2\theta\sin^2\phi }{2(1-\cos\theta  \sin\phi)} + O(\omega^2).
\end{align}
Note that $Q^{(\sfrac{1}{2})}_m$ depends on both the postselected 
state (via $\phi$) and the initial pointer state (via $\theta$),
as shown in Fig.~\ref{Fmq}. 
We can immediately observe  
$Q^{(\sfrac{1}{2})}_m/Q_0 \le 1$.
The equality is satisfied when $(\phi, \theta) = (\pi/2,0)$ and $ (3\pi/2,\pi)$. 
This implies that modular-value-based measurements, like weak-value-based ones
\cite{Alves95,Zhang114,Alves91,Tanaka88,Knee87,Knee4}, cannot 
overcome the standard quantum limit. 
There is no advantage of the modular-value amplification in the qubit-pointer case.    

\begin{figure} [t]
\includegraphics[width=7.0cm]{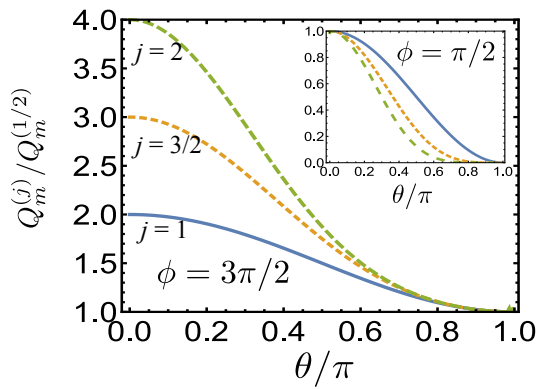}
\caption{
(Color online) $Q^{(j)}_m/Q^{(\sfrac{1}{2})}_m$
for $j =1, 3/2$ and 2 at $\phi = 3\pi/2$ where $|(\sigma_z)_{\rm m}|$ is maximum. 
Its maximum is $2j$ and can be achieved at $\theta =  0$. 
Inset: the same as the main figure at $\phi = \pi/2$ 
where $|(\sigma_z)_{\rm m}|$ is minimum.
}
\label{fig2}
\end{figure}

{\em Higher-dimensional pointer.}---We consider  $Q^{(j)}_m$ for $j\ge1$. 
Detailed calculations are shown in the Supplementary Material \cite{Sub}. 
We obtain $Q^{(j)}_m/Q^{(\sfrac{1}{2})}_m \le 2j$. 
The equality is satisfied when $(\phi, \theta) = (3 \pi/2, 0)$. 
In Fig.~\ref{fig2}, $Q^{(j)}_m/Q^{(\sfrac{1}{2})}_m$ 
is shown as a function of $\theta$ for $j=1, 3/2,$ and 2. 
When $\phi =3\pi/2$ ($|(\sigma_z)_{\rm m}|$ is maximum),  
$Q^{(j)}_m/Q^{(\sfrac{1}{2})}_m$ can be larger than 1. 
On the other hand, its maximum is 1 when $\phi = \pi/2$ ($|(\sigma_z)_m|$ is minimum). 
There is an advantage of the modular-value amplification in the case of $j\ge1$.

Suppose that we have $L+1$ qubits as a resource for measurement. 
We may employ them as $L$ independent 
sensors and one qubit pointer. In this case, $\delta \omega$ scales 
$1/\sqrt{L}$ which corresponds to the standard quantum limit. In order to
measure all sensors, we have to repeat measurements $L$ times.
On the other hand, we can employ them as one qubit sensor 
and a $j \ge 1$  pointer formed with $L$ qubits. Note that 
$j = L/2$ \cite{Giraud114}. Therefore, we expect the enhancement of 
the Fisher information as $Q^{(j)}_m/Q^{(\sfrac{1}{2})}_m = L $ 
according to the above discussion. If we are allowed to measure $L$ times 
as in the previous case, we may be able to expect another $L$ enhancement 
in the Fisher information. In total, the enhancement of the Fisher 
information can scale $L^2$. It implies $\delta \omega$ scales 
$1/{L}$. We claim that
our modular-value-based measurement with a spin-$j$ coherent pointer 
can approach the Heisenberg limit. 

{\em Measurements under noise.}---So far, we assumed the 
ideal noiseless environments. Let us consider measurements under 
noisy environments where phase-flip errors occur on the pointer.
The influence of the noise is described by the operator-sum 
representation as \cite{Nielsen,Desurvire},
\begin{align}\label{rho_noise}
\rho' =(1-\nu)\rho
+\nu{\sigma}_z\rho{\sigma}_z,
\end{align}
where $\rho'$ is the density matrix of the pointer 
under the noise,  $\rho$ is that in the noiseless environment, and 
$0< \nu <1 $ is the probability of the phase-flip. 
When $\nu = 1/2$, we totally lose the information of $\omega$  
\cite{Desurvire}. Hereafter, we only discuss a modular-value-based 
measurement with a qubit-pointer ($j=1/2$) 
as a concrete example of noisy measurements. 

By using the symmetric logarithmic derivative (SLD) operators 
defined by  
$\hat{L}_\lambda\rho+\rho \hat{L}_\lambda 
= 2\partial_\lambda\rho$ \cite{Braunstein72,Paris7}, 
the quantum Fisher information matrix $(H_{\alpha, \beta})$  
is defined by 
\cite{Paris7,Fujiwara201,Matsumoto35,Helstrom25}
\begin{align}\label{Hmatrix}
H_{\alpha, \beta} 
= p(\omega){\rm Tr}\Bigl[\rho'\dfrac{\hat{L}_\alpha \hat{L}_\beta
+\hat{L}_\beta \hat{L}_\alpha}{2}\Bigr],
\end{align}
where $\alpha, \beta = \omega$ or $\nu$. $H_{\omega, \omega}$ 
is the quantum Fisher information associated with the estimation 
of $\omega$ and named as ${}^\omega Q^{(\sfrac{1}{2})}_m$,
while $H_{\nu, \nu}$ is that with the estimation of $\nu$ and named 
as ${}^\nu Q^{(\sfrac{1}{2})}_m$.
For $j = 1/2$, the final density matrix of the pointer is given by 
$\rho = |\eta\rangle\langle\eta|$, where $|\eta\rangle$ 
is given by Eq.~\eqref{pointer_final_conv}. Then, we obtain 
\begin{align}\label{Hmatrix_direct}
H = 
\begin{pmatrix}
\frac{(1-2\nu)^2t^2\sin^2\theta\sin^2\phi}{2(1-\sin\theta\sin\phi)} & 0\\
0 & \frac{\sin^2\theta\cos^2\phi}{2(\nu-\nu^2)(1-\cos\theta\sin\phi)}
\end{pmatrix} + O(\omega^2).
\end{align}
We observe ${}^\omega Q^{(\sfrac{1}{2})}_m
= (1-2 \nu)^2 \ Q^{(\sfrac{1}{2})}_m$: This fact 
clearly illustrates that the noise degrades the measurements
and that the $\theta-$ and $\phi-$dependencies are inherited 
from $Q^{(\sfrac{1}{2})}_m$ to 
${}^\omega Q^{(\sfrac{1}{2})}_m$. 
It is in agreement with the general discussion 
under noise \cite{Jordan2}, too. 

\begin{figure} [t]
\includegraphics[width=7.0cm]{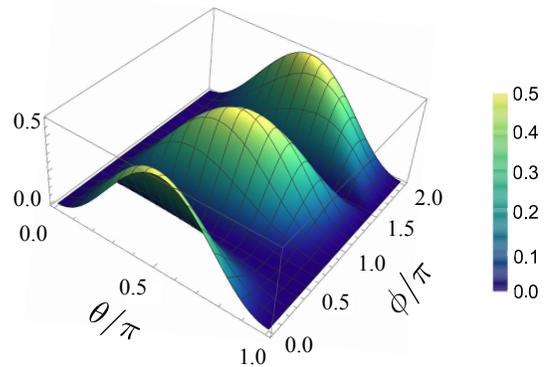}
\caption{
(Color online) 
$ (\nu -\nu^2)\ {}^\nu Q^{(\sfrac{1}{2})}_m$
as a function of $\theta$ and $\phi$. 
}
\label{Fmn}
\end{figure}

We show $ (\nu -\nu^2)\ {}^\nu Q^{(\sfrac{1}{2})}_m$
as a function of $\theta$ and $\phi$ in Fig.~\ref{Fmn}. 
It is interesting to note that  
$ {}^\nu Q^{(\sfrac{1}{2})}_m = 0$ when $\phi = \pi/2$ and 
$3\pi/2$ and that these $\phi$'s give the maxima  of 
$ {}^\omega Q^{(\sfrac{1}{2})}_m$.  
Even more interesting, the parameter combinations of $\theta = \pi/2$ 
and $\phi = n \pi $ ($n$: integer) give the maxima of 
$ {}^\nu Q^{(\sfrac{1}{2})}_m$ at the cost 
of ${}^\omega Q^{(\sfrac{1}{2})}_m= 0$. 
It implies that we can optimize the measurement of $\nu$, or noise, 
by selecting $\theta$ and $\phi$ properly. It is very important 
to measure noise since noise often causes relaxation of quantum systems. 

{\em Conclusions.}---We have investigated the modular-value-based 
measurements with spin-$j$ 
coherent pointers. We discussed two cases with $j = 1/2$ and $j \ge 1$. 
From the viewpoint of quantum Fisher information, 
we first showed the modular-value-based measurements with a qubit pointer
($j=1/2$) has no merit in the sensitivity enhancement as in 
a weak-value-based measurement with a zero-mean Gaussian pointer. 
In contrast, the measurements with $j \ge 1$, 
the quantum Fisher information can become $2j$ times that of the qubit 
pointer. If $L$ qubits consist a spin coherent pointer, the 
Fisher information in one measurement scales as $ L$. 
By taking into account the time required 
for measurements, we claimed that the total Fisher information scales 
as $L^2$ and thus we have a chance to approach the Heisenberg limit. 
Our study in the presence of phase-flip errors shows that 
we can optimize a measurement for detecting noise. 

{\em Acknowledgments.}---This work was supported by CREST(JPMJCR1774), JST.

\bibliography{basename of .bib file}

\clearpage
\newpage

\onecolumngrid

{\bf Supplementary Material for 
``Modular-value-based metrology with spin coherent pointers''}

\appendix
\setcounter{equation}{0}
\section{Quantum Fisher information in the case of 
$j=1/2$: $Q_m^{(1/2)}$}\label{appA}
We show the detailed calculation of the measured quantum 
Fisher information in the case of modular-value-based measurement 
with the spin-$j$ coherent state when $j = 1/2$. Let remind us  
the final pointer state
\begin{align}\label{final_ps_spin1/2_app}
|\eta\rangle = \dfrac{1}{\sqrt{2}}\Bigl(\cos\frac{\phi}{2}|\xi_
+\rangle+\sin\frac{\phi}{2}e^{-i\omega t}|\xi_-\rangle\Bigr)/\sqrt{p(\omega)},
\end{align}
where, 
\begin{align}
&|\xi_\pm\rangle  = e^{-i\varphi}\sin\frac{\theta}{2}
\Bigl|\frac{1}{2},-\frac{1}{2}\Bigr\rangle \mp i \cos\frac{\theta}{2}
\Bigl|\frac{1}{2},+\frac{1}{2}\Bigr\rangle, \label{xipmspin1/2_app} \\
&p(\omega)  = (1 - \cos\theta  \sin\phi \cos \omega t)/2, \label{pomg_app}\\
& \langle\xi_\pm|\xi_\pm\rangle=1, \ \langle\xi_-|\xi_+\rangle 
= \langle\xi_+|\xi_-\rangle=-\cos{\theta}.\label{normorth_app}
\end{align}
Taking the derivative of the final pointer state, we obtain
\begin{align}\label{dnu}
\dfrac{d |\eta\rangle}{d\omega}  = \dfrac{-it}{\sqrt{2p(\omega)}}
\Bigl(\sin\frac{\phi}{2}e^{-i\omega t}|\xi_-\rangle\Bigr).
\end{align}
Then, we have
\begin{align}\label{dnu}
\dfrac{d \langle\eta|}{d\omega}\dfrac{d |\eta\rangle}{d\omega}  
&= \dfrac{t^2}{2p(\omega)}\sin^2\frac{\phi}{2}, \ \text{and }\\
\dfrac{d \langle\eta|} {d\omega}|\eta\rangle 
&= \dfrac{it}{2p(\omega)}\Bigl[\sin^2\frac{\phi}{2}
-\frac{1}{2}\cos\theta\sin\phi e^{i\omega t}\Bigr].
\end{align}
And thus, the measured quantum Fisher information yields 
\begin{align}\label{fisher_modular_app}
\notag Q_m^{(\sfrac{1}{2})} &= 4p(\omega)
\Bigl[\dfrac{d \langle\eta|}{d\omega}
\dfrac{d |\eta\rangle}{d\omega} -\Bigl|\dfrac{d \langle\eta|} 
{d\omega}|\eta\rangle\Bigr|^2\Bigr]\\
&=\dfrac{t^2\sin^2\theta\sin^2\phi }{2(1-\cos\theta  \sin\phi)} 
+ O(\omega^2).
\end{align}
Equation~\eqref{fisher_modular} in the main text is obtained. 

To calculate the postselected classical Fisher information 
$F_p$, we substitute
\begin{align}\label{post_Fish_app}
\dfrac{dp(\omega)}{d\omega}=\dfrac{t}{2}\cos\theta\sin\phi\sin\omega t 
\end{align}
to Eq.~\eqref{Fisher_p}, then we obtain $F_p = O(\omega^2)$.

\section{Quantum Fisher information in the case of 
$j\ge 1$: $Q_m^{(j)}$}\label{appB}
The final pointer state is the same 
as Eq.~\eqref{final_ps_spin1/2_app}, where
\begin{align}\label{xipmspinj>1_app}
\notag|\xi_\pm\rangle &= \exp\Bigl(\mp i\frac{\pi}{2}|j,+j\rangle
\langle j,+j|\Bigr)|\xi\rangle\\
 &=\sum_{m=-j, j\ge1}^{j-1}c_{jm}|j,m\rangle \mp i 
\Bigl(\cos\frac{\theta}{2}\Bigr)^{2j}|j,+j\rangle,
\end{align}
and 
\begin{align}\label{Sigma_app}
c_{jm}= \sqrt{{2j \choose j+m}}\Bigl(\cos\frac{\theta}{2}
\Bigr)^{j+m}\Bigl(\sin\frac{\theta}{2}e^{-i\varphi}\Bigr)^{j-m},\ -j\le m\le j-1.
\end{align}
Then, we obtain 
\begin{align}\label{rela_app}
 \langle\xi_\pm|\xi_\pm\rangle=1,\ \text{and } 
\langle\xi_-|\xi_+\rangle = \langle\xi_+|\xi_-\rangle
=\sum_{m=-j}^{j-1}c_m^2-(\cos\frac{\theta}{2})^{4j},
 \end{align}
 as in the main text. Similar to the case of $j = 1/2$, 
we calculate the derivative of the final pointer state as
\begin{align}\label{dnu_appB}
\dfrac{d |\eta\rangle}{d\omega}  
= \dfrac{-it}{\sqrt{2p(\omega)}}\Bigl(\sin\frac{\phi}{2}
e^{-i\omega t}|\xi_-\rangle\Bigr).
\end{align}
Then, we have
\begin{align}\label{dnu_appB}
\dfrac{d \langle\eta|}{d\omega}\dfrac{d |\eta\rangle}{d\omega}  
= \dfrac{t^2}{2p(\omega)}\sin^2\frac{\phi}{2}, \ \text{and } 
\dfrac{d \langle\eta|} {d\omega}|\eta\rangle 
= \dfrac{it}{2p(\omega)}\Bigl\{\sin^2\frac{\phi}{2}
+\frac{1}{2}\Bigl[\sum_{m=-j}^{j-1}c_m^2-(\cos\frac{\theta}{2})^{4j}\Bigr]
\sin\phi e^{i\omega t}\Bigr\}.
\end{align}
The measured quantum Fisher information yields 
\begin{align}\label{fisher_modular_app}
\notag Q_m^{(j)} 
&= 4p(\omega)\Bigl[\dfrac{d \langle\eta|}{d\omega}
\dfrac{d |\eta\rangle}{d\omega} -\Bigl|\dfrac{d \langle\eta|} 
{d\omega}|\eta\rangle\Bigr|^2\Bigr]\\
&=t^2\Biggl\{2\sin^2\dfrac{\phi}{2}-\dfrac{1}
{p(\omega)}\Biggl[\sin^2\frac{\phi}{2}+\frac{1}{2}
\Bigl(\sum_{m=-j}^{j-1}c_m^2-(\cos\frac{\theta}{2})^{4j}\Bigr)
\sin\phi e^{i\omega t}\Biggr]^2\Biggr\} + O(\omega^2),
\end{align}
where 
\begin{align}\label{poappB}
p(\omega)=\dfrac{1}{2}
\Bigl[1+\Bigl(\sum_{m=-j}^{j-1}c_m^2-(\cos\frac{\theta}{2})^{4j}\Bigr)
\cos\phi\cos\omega t\Bigr].
\end{align}
From this result, we can easily calculate the measured quantum Fisher 
information for any given $j$.

\end{document}